# A code-free optical undersampling technique for broadband microwave spectrum measurement

Guangyu Gao
*Qian Xuesen Laboratory of Space Technology*
*China Academy of Space Technology*
Beijing, China
gaoguangyu@qxslab.cn

Xueshuang Xiang
*Qian Xuesen Laboratory of Space Technology*
*China Academy of Space Technology*
Beijing, China
xiangxueshuang@qxslab.cn

Qijun Liang
*Qian Xuesen Laboratory of Space Technology*
*China Academy of Space Technology*
Beijing, China
liangqijun@qxslab.cn

Naijin Liu
*Qian Xuesen Laboratory of Space Technology*
*China Academy of Space Technology*
Beijing, China
liunaijin@qxslab.cn

*Abstract*—A novel broadband microwave (MW) spectrum measurement (BMSM) scheme based on code-free optical under-sampling and homodyne detection is proposed. The fully analog generation of optical pulses with a far-less-than-Nyquist rate is only through modulating cascaded electrooptical modulators by a single RF tone instead of any high-speed coding sequence modulation. Homodyne detection will reduce the analysis bandwidth of BMSM and enhance the detection performance of weak signal. A multi-band signal with 20 GHz spectral range and SNR = 61 dB is used to investigate the BMSM performance of this scheme, and the results show good performance for BMSM. The potentials for further optimization in practice are also discussed.

## I. Introduction

Photonic techniques has been introduced into BMSM for various application Scenarios in the past decades as next-generation solutions to overcome the limitations of traditional all-electric schemes, due to the prominent advantages of photonic techniques such as broad bandwidth, low loss, high-efficiency and tunable filtering and local oscillator (LO) and intrinsic immunity to electromagnetic interference [1-6]. Among different photonics-based BMSM schemes, optical undersampling techniques as powerful solutions can sense broadband MW signal sparse in frequency domain at ADC sampling rate far less than Nyquist rate with relatively simple architectures. Recently, several photonics-based compressive sampling (CS) schemes [5-9] as a typical kind of under-sampling techniques exhibit great advantages in reducing analysis bandwidth and data amount. Although the ADC sampling rate in these schemes are largely reduced, high-speed optical pulses with coding sequence modulation (e.g. pseudo-random bit sequence, PRBS) at or above Nyquist rate are still required for incoherence sampling and random demodulation, which are the major bottleneck in hardware implementation for these schemes. Multi-channel methods in [6-8] have been employed to achieve higher operation band-widths and lower analysis bandwidths, but high-speed coding sequence and accurate delay control among channels are still required, leading to additional complications and penalties.

In this paper, we propose a novel BMSM scheme based on code-free optical undersampling technique. An OFC unit generates code-free optical pulses for undersampling without any coding sequence modulation. The repetition rate of the optical sampling pulse could be much lower than the Nyquist rate as compared with traditional schemes. The homodyne detection can extract both in-phase (I) and quadrature (Q) components of the optical sampling signal, making the measurement values in positive and negative frequency ranges both available and great enhancing the detection performance of weak signal. After low-pass filtering and quantization by low-speed electrical ADC, the down-converted signal is fed into the digital signal processing (DSP) unit for spectrum reconstruction of the broadband MW signal.

## II. Principle

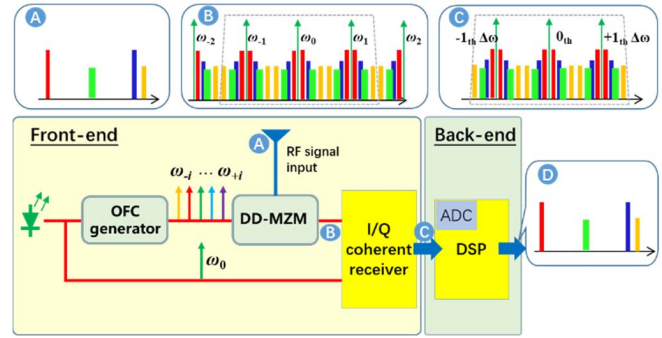

Figure 1 the schematic of the proposed photonics-based BRSM system

The proposed scheme is schematically illustrated in Fig. 1, comprised of two parts, the optical front-end and the electrical back-end. In the front end, a continuous wave (CW) light with frequency at ω0 from a narrow linewidth laser is divided into two paths by a power splitter, one path is sent to the OFC generator, and the other as a local oscillator (LO) is sent to the I/Q coherent receiver for homodyne detection. In the OFC generator, the CW light is electro-optic modulated to generate an OFC with cascaded intensity and phase modulators driven by a RF tone at frequency Δω. The generated OFC are then intensity modulated with an intensity modulator by a broad-band MW signal under detection as shown in fig 1A. Upon intensity modulation, all frequency components of the MW signal are up-converted on light as shown in fig 1B. these process is also named as optical sampling, which is expressed in time domain as $y(t) = p(t) \times x(t)$, where $x(t)$ and $y(t)$ are the real MW signal before and after optical sampling. $p(t)$ is the optical sampling pulses given by $\exp[i(\omega_0 t + \beta_0 \cos(2\pi\Delta f t) + \pi/4)] \times \cos(\beta_1 \cos(2\pi\Delta f t) + \pi/4)$, where $\beta_0$ and $\beta_1$ are respectively the modulation factors of the phase and intensity modulators. Through mixed with the LO and homodyne detection, the sampled signal is down-converted into electrical domain and fed into the back end for digital signal processing for signal reconstruction. The spectrum of the down-converted signal is expressed as

$$Y = 2A_{LO}A_S \sum_{k=-K}^{K} S_k \sum_{i=0}^{I} \left( X(f_i) + X(f_i)^* \right) + N(f). \quad (1)$$

where $Y$ is the spectrum of the down-converted signal of $y(t)$, $\sum_0^H (X(f_h) + X(f_h)^*)$ is the spectrum of $x(t)$, and $S_k$ is the kth comb line of the OFC. $N(f)$ is the noise component.

For a frequency component $f + l\Delta\omega$ in the down-converted signal, its measured values $Y_l (l\Delta\omega + f)$ and the symmetric component $Y_l (l\Delta\omega - f)$ can be written as

$$Y_l(f + l\Delta f) + Y_l(l\Delta f - f) = \sum_{j=-J}^{+J} S_k \times \{X(f + j\Delta f) + X(j\Delta f - f)\} + N(l\Delta f \pm f). \quad (2)$$

where $k = l - j$, $L = K + J$, L and J $\in \mathbb{N}$, $K \geq J, j \in [-J, J]$, $l \in [-L, L]$; and for $|k| > K$, $S_k = 0$; for $|k| \leq K$, $S_k \neq 0$.

According to equation (1) and (2), the components of MW signal belonging to the set $\mathcal{F} = \{l\Delta\omega \pm f\}$ will subject to frequency aliasing, needing suitable algorithm to reconstruct the original components. Different from the traditional schemes in which only the measurement values in a single side band are available by direct detection of optical sampling signal, the down-converted information in dual side bands are both extracted under homodyne detection as shown in fig 1C, doubling the number of the measurement values. For an undersampling system, the down-converted signal will go through a low-pass filter (LPF) with a cut-off frequency far less than the Nyquist rate of the MW signal to reduce electric sampling rate and the amounts of quantization data. Based on the assumption that the frequency response of the LPF is an ideal rectangular function, the spectrum of the filtered signal with components belonging to the set $\mathcal{F}^+ = \{l\Delta\omega + f\}$ can be given in matrix form $Y = AX$, expressed as

$$\begin{bmatrix} Y_{-L} \\ \vdots \\ Y_l \\ \vdots \\ Y_{L-1} \end{bmatrix} = \begin{bmatrix} S_{-L} & S_{-L-1} & \cdots & S_{-L-2J+1} & S_{-L-2J} \\ \vdots & \vdots & \vdots & \vdots & \vdots \\ S_k & S_{k-1} & \vdots & S_{k-2J+1} & S_{k-2J} \\ \vdots & \vdots & \vdots & \vdots & \vdots \\ S_{L-1} & S_{L-2} & \cdots & S_{L-2J} & S_{L-2J-1} \end{bmatrix} \begin{bmatrix} M_{f-J\Delta\omega} \\ \vdots \\ M_f \\ \vdots \\ M_{f+J\Delta\omega} \end{bmatrix} \quad (3)$$

where $A$ is a $2L \times (2J+1)$ measurement matrix with $L \ll J$, $k$ and $l \in [-L, L-1]$.

By combining equation 3 and the matrix form of the symmetric component $\mathcal{F}^- = \{l\Delta\omega - f\}$, a combination equation can be constructed with double measurement values over equation (3). Equation (3) and its combination equation are underdetermined and cannot be solved directly. While it is possible to find the unique sparsest solution or some approximate sparse ones of $X(f,j)$ by given some additional criteria [10-11], provided the original spectrum is sparse to some extent. As for BMSM scenario, the nonzero-value distribution of the broadband MW spectrum in practice is highly sparse, normally described by multiband models [12-13]. Thus, by adopting a suitable algorithm based on a basic criterion of L0–norm optimization expressed in equation (4), one can reconstruct the spectrum of an unknown broadband MW signal in a high fidelity.

$$\arg\min \|X\|_0 \quad \text{s.t.} \quad Y = AX. \quad (4)$$

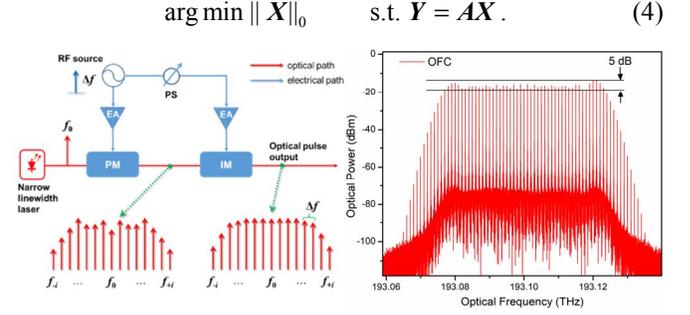

Fig. 2. (a) The schematic of the electro-optic-based OFC generator in front end, and (b) the optical spectrum of OFC; PS: phase shifter; EA: electrical amplifier; PM: phase modulator; IM: intensity modulator.

### III. RESULTS AND DISCUSSION

The proposed scheme as shown in fig. 1 is demonstrated by computer simulation. The power, frequency and linewidth of the CW light is set at 16 dBm, 193.1 THz and 100 kHz. In the OFC generator, the phase modulator and intensity modulator biased at quadrature point are modulated with a RF tone at 1 GHz to achieve an OFC with spacing at 1 GHz, corresponding to an optical pulse sequence with far-less-than-Nyquist rate at 1 GHz. The phase modulator is set with modulation factor $\beta_0$ at 7.25 $\pi$, and the intensity modulator biased at quadrature point is set with parameters of half-wave voltage at 4 V, modulation factor $\beta_1$ at 0.3 $\pi$, insertion loss at 5 dB, and extinction ratio at 30 dB. The phase difference between two modulators are 0.1 rad. By well setting the driving voltages and phase difference between the modulators, an OFC is generated as shown in Fig. 2a, which has 47 coherent comb lines from − 23th to + 23th with flatness less than 5 dB. The generated OFC is directly used to sample the MW signal without further spectral optimization. The bandwidths of the photodiodes (PD) is set at 2.5 GHz. The down-converted signal is further filtered and sampled by a low-pass filter with cut-off frequency at 2 GHz and an ADC with sampling rates equal to 4 GSa/s, which gives a compressive ratio at 10 due to the Nyquist rate of the MW signal equal to 40 GHz. The frequency resolution is 1.22 MHz.

The performance of the proposed scheme was investigated with a multi-band signal generated by a Matlab module. Its spectral range is from 0 to 20 GHz as shown in fig 3a, containing three wideband NRZ signals with carrier frequencies and bandwidths respectively at 7.52 GHz (100 MHz), 10.25 GHz (50 MHz), 19.7 GHz (30 MHz) and white noise with power spectral density at -146 dBm/Hz. The signal-to-noise ratio (SNR) of the MW signal is 61 dB. The spectrum of the down-converted signal is shown in fig 3b. 8 measurement values within the range from

-2GHz to +2GHz are used to reconstruct the signal spectrum. The comb lines of the OFC used to construct the matrix $A$ are selected within the spectral range as shown in fig 2a from 193.079 THz to 193.121 THz ($k$ = - 21 to + 21), coverring a single-side-band range at 21 GHz far less than the Nyquist rate at 40 GHz.

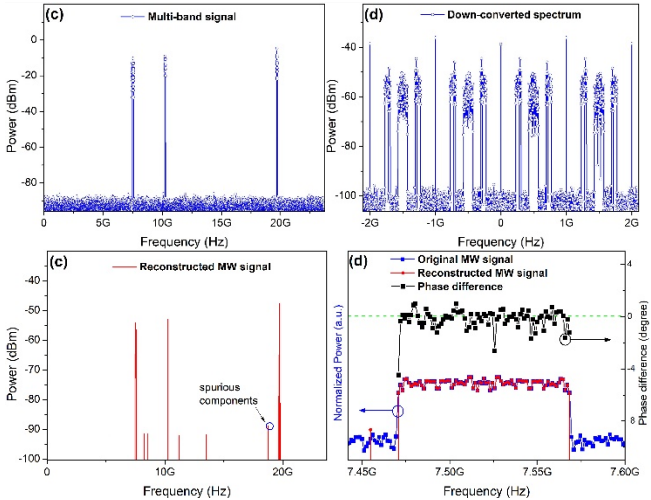

Figure 3 (a) the spectrum of the MW signal under detection; (b) the spectrum of the down-converted signal; (c) the spectrum of the reconstructed signal; (d) the spectrum comparison (lower curves) and the phase difference (upper curve) between the original wideband signal with carrier frequency at 7.52 GHz and its reconstructed signal

The reconstruction results are shown in fig 3c and 3d. The main frequency components of the MW signal with powers greater than the noise floor $T$ at -88 dBm were reconstructed evidently as shown in fig 3c, and most of noise components were eliminated except for several small spurious signals with powers slightly above $T$. The spectrum comparison and phase difference between the original wideband signal with carrier frequency at 7.52 GHz and its reconstructed signal are shown in fig 3d in detail. The relative reconstruction error $E_r$ is 0.004 for the down-converted signal with noise floor at -96 dBm as shown in fig 3b. $E_r$ is given by $\|X-\hat{X}\|_2/\|X\|_2$, where $\hat{X}$ is the reconstructed coefficient vector of $X$. The detection limit of the MW signal in SNR is 6 dB, and the spurious free dynamic range (SFDR) is 59 dB (99 dB·Hz$^{2/3}$) not considering the spurious signal from reconstruction error. The results above indicate that the proposed scheme is feasible for BMSM. Furthermore, this scheme possesses potentials for further performance optimization and system development in practice, for example, the optimization of the measurement matrix based on the configuration of OFC, the increase of detection bandwidth and the increase of the reconstruction efficiency.

## IV. CONCLUSION

In conclusion, we have proposed and demonstrated a BMSM scheme based on code-free optical undersampling and homodyne detection, achieving high-fidelity spectrum reconstruction of an unknown broadband sparse MW signal. The distinct advantages of this scheme are threefold. Firstly, the optical sampling pulse with repetition rate far less than the Nyquist rate is generated only by modulating a CW laser with a low-frequency RF tone without any high-speed coding sequence modulation such as high-bit-rate PRBS, breaking through the bottlenecks of traditional schemes on the requirements for high-bit-rate electrical devices. Secondly, the repetition rate, optical spectral range and characteristics of OFC spectrum are easy to be reconfigured based on various requirements. Thirdly, the homodyne detection of optical sampling signal reduces the analysis bandwidth of BMSM and significantly enhance the detection performance of weak signal.

ACKNOWLEDGMENT

This work was supported by Independent Innovation Fund of Qian Xuesen Laboratory of Space Technology, and Independent research and development projects of China Aerospace Science and Technology Corporation.